\documentclass[aps,prd,twocolumn,showpacs,showkeys,preprintnumbers]{revtex4}
\usepackage{epsfig}
\usepackage{amssymb}
\newcommand{\beq}{\begin{equation}}
\newcommand{\eeq}{\end{equation}}
\newcommand{\beqn}{\begin{eqnarray}}
\newcommand{\eeqn}{\end{eqnarray}}
\newcommand{\nn}{\nonumber}
\newcommand{\ep}{\epsilon}
\newcommand\as{{\alpha_s}}

\def\bom#1{{\mbox{\boldmath $\mathrm{#1}$}}}
\def\res#1#2#3{$(#1\pm#2)\cdot 10^{#3}$}

\begin{document}

\preprint{hep-ph/0307268}
\title{Next-to-leading order calculation of three-jet observables
  in hadron-hadron collision}
\author{Zolt\'an Nagy}
\email[]{nagyz@physics.uoregon.edu}
\homepage[]{http://www.cpt.dur.ac.uk/~nagyz/nlo++}
\affiliation{Institute of Theoretical Science,
5203 University of Oregon,
Eugene, OR  97403-5203, USA
}

\date{\today}

\begin{abstract}
The production of the three jets in hadron-hardon collision is the
first more complex process which allow us to define a branch of
variables in order to do more precise measurement of the strong
coupling and the parton distribution function of the proton. This
process is also suitable for studying the geometrical properties of
the hadronic final state at hadron colliders. This requires next-to-leading
order prediction of the  three-jet observables.
In this paper we describe the theoretical formalism of such a
calculation with sufficient details. We use a the dipole method to
construct Monte Carlo program for calculating three-jet observables at
next-to-leading order accuracy. We present a theoretical prediction
for inclusive and exclusive cross section and for some relevant event
shape variables like transverse thrust, transverse jet broadening  and
$E_{t3}$ variable.  
\end{abstract}

\maketitle

\section{Introduction}

In high energy hadron-hadron collision the processes with purely
hadronic final state is one of the clearest processes to test the
Quantum Chromodynamics (QCD) an measure its parameters simultaneously.
According to the perturbative QCD the cross section of this processes 
is a convolution of a long- and a short distance part. 

The long distance dependent part is the parton
distribution function of the incoming hadrons. This function is
universal, process independent and we can measure it in any basics
process for example deeply inelastic scattering.  

The short distance part is the partonic cross section that can be
calculated in perturbative QCD as a function of a single parameter,
the strong coupling ($\as$). The main advantage of the hadronic final 
states at hadron-hadron colliders is that it is possible to measure
the strong coupling and the PDF function simultaneously.   

The production of three jets is the first process which can provide
complex final state to define a branch of jet observables in
order to be able to do more precise measurement of the strong coupling
and give better determination of the parton distribution function. On
the other hand this process allow us to do more advanced studies of
the hadronic final states by measuring its geometrical properties. 
In order to be able to make quantitative prediction it is essential to
perform the computations at least next-to-leading (NLO) accuracy.  In
hadron collision the most easily calculated one- and two-jet cross
sections have so far been calculated at NLO level \cite{jetrad, ks}.
At next-to-leading level some inclusive three-jet observables were
calculated by Giele and Kilgore \cite{Kilgore:2000dr,Kilgore:1997sq}
and by the author \cite{Nagy:2001fj}. Furthermore, Z. Tr\'ocs\'anyi
also calculated the three-jet cross section in effective parton
distribution function approximation \cite{Trocsanyi:1996by}.

The main difficulty of the next-to-leading order calculations is the
presence of the singularities. In general, when evaluating higher
order QCD cross section, one has to consider real-emission
contributions and virtual corrections and one has to deal with
different type of the singularities. The ultraviolet singularities
presented in the virtual contributions and they are removed by the
renormalization. The infrared singularities presented both in the
virtual and real contribution but the sum of them is finite.     
In the last few years the theoretical developments made possible the
next-to-leading order calculation for three-jet quantities. There
are several general methods available for the cancellation of the
infrared divergences that can be used for setting up a Monte Carlo
program \cite{Nagy:1997bz,Frixione:1996ms,Catani:1997vz}. In computing
the NLO correction we use the dipole formalism of Catani-Seymour
\cite{Catani:1997vz} that we modified slightly in order to have a
better control on the numerical calculation. This scheme is discussed
in details in the next section.

The advantages of using the dipole method are the followings: i) no
approximation is made; ii) the exact phase space factorization allows
full control over the efficient generation of the phase space; iii)
neither the use of color ordered subamplitudes, nor symmetrization,
nor partial fractioning of the matrix elements is required; iv)
Lorentz invariance is maintained, therefore, the switch between
various frames can be achieved by simply transforming the momenta; v)
the use of crossing functions is avoided; vi) it can be implemented in
an actual program in a fully process independent way.

The important theoretical development that made possible the three-jet
calculation was that the relevant one-loop amplitudes for the relevant
subprocesses became available. For the $0\to ggggg$ \cite{amp-g5},
$0\to q\bar{q}ggg$ subprocesses the amplitudes were calculated by
Bern, Dixon and Kosower \cite{amp-q2g3} and for the $0\to
q\bar{q}Q\bar{Q}g$ subprocess it was given by Kunszt, Signer and
Tr\'ocs\'anyi \cite{amp-q4g1}.  The relevant six parton tree-level
matrix elements are also available
\cite{Gunion:1985bp,Gunion:1986zh,Gunion:1986zg,Kuijf:1991kn}.

In this paper we give sufficient details of our work and present
several new results for next-to-leading order prediction of three-jet
observables in hadron-hadron collision that were not published before
and they could be interesting in an experimental analysis.
In the Sec.~\ref{sec:method} we give details of the analytic and
numeric calculation and describe the structure of our result. In the
Sec.~\ref{sec:result} we present complete NLO predictions for
three-jet inclusive cross sections and their energy fraction
distribution (Dalitz variables) using the inclusive $k_\perp$-algorithm
\cite{Ellis:1993tq} and the midcone algorithm \cite{Blazey:2000qt}. We
define the thrust and jet broadening event shapes variables in the
transverse plane. We show some subjet result using the exclusive
$k_\perp$-algorithm \cite{Catani:1993hr}. Sec.~\ref{sec:conclusion}
contains our conclusion.

\section{\label{sec:method}Details of the calculation}

\subsection{The method}

The next-to-leading order cross section for the process with two 
initial state hadrons is the convolution of the parton density 
functions of the incoming hadrons and the hard scattering cross section, 
\beqn
\label{had-xsec}
\nn
\sigma(p, \bar{p}) = \sum_{a,b}
\int_0^1\!\!d\eta \, f_{a/H}(\eta,\mu_F^2)
\int_0^1\!\!d\bar{\eta}\,f_{b/\bar{H}}(\bar{\eta},\mu_F^2)\\
\times\left[\sigma^{LO}_{ab}(\eta p, \bar{\eta}\bar{p}) 
  + \sigma^{NLO}_{ab}(\eta p, \bar{\eta}\bar{p})\right]\;\;,
\eeqn
where $f_{i/H}$ is the density of the partons of type $i$ in the incoming
hadron $H$ at the  $\mu_F$ factorization scale. The corresponding parton level 
cross sections are
\beqn
\sigma^{LO}_{ab}(p, \bar{p}) \equiv \int_3 d\sigma^{B}_{ab}(p, \bar{p}) 
= \int_3 d\Gamma^{(3)}\langle |{\cal M}^{(3)}_{ab}|^2\rangle F_J^{(3)}\;\;,
\eeqn
and the next-to-leading order correction is sum of three integrals
\beqn
\label{par-xsec-nlo}
\nn
&&\sigma^{NLO}_{ab}(p, \bar{p}) \equiv \int d\sigma^{NLO}_{ab}(p, \bar{p})\\
&&\quad = \int_4 d\sigma^{R}_{ab}(p, \bar{p}) + \int_3 d\sigma^{V}_{ab}(p, \bar{p})
 + \int_3 d\sigma^{C}_{ab}(p, \bar{p})\;\;,\qquad
\eeqn
where $d\sigma^{R}$ and $d\sigma^{V}$ are the real and virtual contributions 
to the cross section. The contribution $d\sigma^{C}$ represents the collinear 
subtraction counter term. The pole structure of this term is well
defined but the finite part is factorization scale dependent. The actual 
form of the finite part in the expression of the $d\sigma^{C}$ defines the
factorization scheme. In our calculations we chose the commonly used 
$\overline{\rm MS}$ subtraction scheme. The parton density functions are also 
scheme dependent, so that this dependence cancels in the hadronic cross 
section of Eq. (\ref{had-xsec}).

The three integrals on the right hand side of Eq. (\ref{par-xsec-nlo}) are 
separately divergent but their sum is finite provided by the jet function
$F_J^{(m)}$ defines an infrared safe quantity, which formally means that 
\beq
F_J^{(4)} \longrightarrow F_J^{(3)}\;\;,
\eeq
in any case where the four-parton and three-parton configurations are 
kinematically degenerate. The presence of the singularities means that the 
separate pieces have to be regularized and the divergences have to be 
cancelled. We use the dimensional regularization in $d=4-2\epsilon$ dimensions
in which case the divergences are replaced double poles $1/\epsilon^2$ and 
single poles $1/\epsilon$. We assume that ultraviolet renormalization of all 
Green functions to one-loop order has been carried out, so the divergences are 
infrared origin. In order to get the finite sum a slightly modified version of
the Catani-Seymour \cite{Catani:1997vz} dipole method is used in our 
calculations. 

The reason for modifying the original dipole formalism is numerical. The essence 
of the dipole method is to define a single subtraction term $d\sigma^{A}$, 
the dipole subtraction term, that regularizes the divergences in the real 
term which comes form the soft and collinear regions. Thus, the three singular
integrals in Eq. (\ref{par-xsec-nlo}) are substituted by three finite ones:
\beqn
\label{final-xsec}
\nn
&&\sigma^{NLO}_{ab}(p, \bar{p}) = \sigma^{NLO}_{ab\{4\}}(p, \bar{p})
+ \sigma^{NLO}_{ab\{3\}}(p, \bar{p}) \\
&&\quad+ \int_0^1 dx \left[\hat{\sigma}^{NLO}_{ab\{3\}}(x, xp, \bar{p})+
  \hat{\sigma}^{NLO}_{ab\{3\}}(x, p, x\bar{p})\right]\;\;, \qquad
\eeqn
where the four-parton integral is given by
\beqn
\sigma^{NLO}_{ab\{3\}}(p, \bar{p}) = \int_4
\left[d\sigma^{R}_{ab}(p, \bar{p})_{\epsilon=0}
  - d\sigma^{A}_{ab}(p, \bar{p})_{\epsilon=0}\right]\;\;.
\eeqn
We have two three-parton contribution to the NLO correction. The second term 
on the right hand side of Eq. (\ref{final-xsec}) is the sum of the one-loop 
contribution and a Born term convoluted by an universal singular factor 
$\bom{I}$,
\beqn
\sigma^{NLO}_{ab\{3\}}(p, \bar{p}) = \int_3
\left[d\sigma^{V}_{ab}(p, \bar{p}) + d\sigma^{B}_{ab}(p, \bar{p})\otimes 
  \bom{I}\right]_{\epsilon=0}\;\;.
\eeqn
The factor $\bom{I}$ contains all the $\epsilon$ poles which come from the 
$d\sigma^{A}$ and  $d\sigma^{C}$ that are necessary to cancel the (equal 
and with opposite sign) poles in $d\sigma^{V}$. The last term in 
Eq. (\ref{final-xsec}) is a finite remainder that is left after factorization of initial-state collinear singularities into the non-perturbative distribution functions (parton density function),
\beqn\nn
&&\int_0^1 dx \, \hat{\sigma}^{NLO}_{ab\{3\}}(x, xp, \bar{p}) =
 \sum_{a'}\int_0^1 dx\\
&& \qquad\int_3
\left[d\sigma^{B}_{a'b}(xp, \bar{p})\otimes (\bom{P}(x) + \bom{K}(x))^{aa'}
\right]_{\epsilon=0}\;,
\eeqn
where the $x$-dependent functions $\bom{P}$ and $\bom{K}$ are similar (but finite for 
$\epsilon\to 0$) to the factor $\bom{I}$. These functions are universal, that 
is, they are independent of scattering process and of the jet observables.

There are several ways to define the $d\sigma^{A}$ dipole subtraction term, 
but all must lead to the same finite next-to-leading order correction. In this 
program a slight modification of the Catani-Seymour subtraction
term was implemented by defining the dipole term as a function of a parameter 
$\alpha \in (0,1]$ which controls the volume of the dipole phase
space. In the $e^+e^-$ annihilation case this modification of the dipole
formalism was discussed in Ref.~\cite{Nagy:1998bb} what we generalize
for the hadron-hadron collision.

In this subsection we recall only those dipole factorization formulaes that are 
relevant in our calculation. We don't want to give the precise definition of 
every variables, functions, factors we just use the same notation of the 
original paper of the dipole method \cite{Catani:1997vz}. 

The $d\sigma^A$ local counter term is provided by the dipole factorization of 
the tree level squared matrix element. Thus we can define 
\begin{widetext}
\begin{eqnarray}\nn
  \label{dipole-terms}
  d\sigma_{ab}^A &=& \sum_{\{4\}} d\Gamma^{(4)}(p_a,p_b,p_1,...,p_4)
  \frac1{S_{\{4\}}}\\\nn
  &&\qquad\times\Bigg\{\sum_{\mathrm{pairs}\atop i,j} \sum_{k\neq i,j}
  {\cal D}_{ij,k}(p_a,p_b,p_1,...,p_4)
  F_J^{(3)}(p_a,p_b,p_1,..,\tilde{p}_{ij},\tilde{p}_{k},..,p_4)
  \Theta(y_{ij,k} < \alpha)\\\nn
&&\qquad\qquad+ \sum_{\mathrm{pairs}\atop i,j}
  \left[{\cal D}_{ij}^a(p_a,p_b,p_1,...,p_4)
  F_J^{(3)}(\tilde{p}_a,p_b,p_1,..,\tilde{p}_{ij},..,p_4) 
  \Theta(1-x_{ij,a} < \alpha) + (a\leftrightarrow b)\right]\\\nn
&&\qquad\qquad+ \sum_{i\neq k}\left[
  {\cal D}_k^{ai}(p_a,p_b,p_1,...,p_4)
  F_J^{(3)}(\tilde{p}_a,p_b,p_1,..,\tilde{p}_k,..,p_{4}) 
  \Theta(u_{i} < \alpha) + (a\leftrightarrow b)\right]\\
&&\qquad\qquad+ \sum_{i}\left[
  {\cal D}^{ai,b}(p_a,p_b,p_1,...,p_{4})
  F_J^{(3)}(\tilde{p}_a,p_b,\tilde{p}_1,...,\tilde{p}_{4}) 
  \Theta(\tilde{v}_i < \alpha) + (a\leftrightarrow b)\right]\Bigg\}\;\;,
\end{eqnarray}
\end{widetext}
where $d\Gamma^{(4)}$ is the four-parton phase space including all the
factors that are QCD independent, $\sum_{\{4\}}$ denotes the sum over
all the configurations with $4$ partons and $S_{\{4\}}$ is the Bose
symmetry factor of the identical partons in the final state. The ${\cal
  D}_{ij,k}$, ${\cal D}_{ij}^a$, ${\cal D}_k^{ai}$ and ${\cal
  D}^{ai,b}$ are the dipole terms. The function $F_J^{(3)}$ is the jet
function which acts over the three-parton dipole phase space. The
$y_{ij,k}$, $x_{ij,a}$, $u_{i}$, $\tilde{v}_i$ are the dipole
variables defined by the dipole factorization of the phase space. The
dipole factorization of the phase space is exact phase space
factorization which means there is no approximation used in the
kinematically degenerated regions (soft, collinear and
soft-collinear). We found it makes the Monte Carlo integral more
stable because the real contribution and the subtraction terms are
defined in same phase space point.

In Eq.~(\ref{dipole-terms}) the $\alpha =1$ case means the full dipole 
subtraction and it gives back the original dipole subtraction terms. In a 
computer program the large number of dipoles terms and complicated analytic 
structure of the expressions makes the evaluation of the subtraction terms 
rather time consuming. Using this cut dipole phase space we can speed 
up the program. The parameter $\alpha$ is also useful to check our program by 
varying the value of $\alpha$ and checking whether the full correction is 
independent of this parameter. 

The most serious numerical defect of the subtraction schemes is the
missed binning. This happens when a huge positive weight from the
real part and the corresponding huge negative weight form the
subtraction term are filled into different histogram bins. It is
obvious this cut can increase the numerical stability of the Monte
Carlo program by decreasing the size of the dipole phase space which
reduces the chance of the missed binning. 

The introduced $\alpha$ phase space cut parameter requires to redefine the 
$\bom{I}$ and $\bom{K}(x)$ flavour kernels. The singular factor $\bom{I}$ 
is given by
\beqn\nn
{\bf I}(p_1,...,p_m;\alpha;\ep) &=& -\frac\as{2\pi}\frac1{\Gamma(1-\ep)}
\sum_I
\frac1{\bom{T}_I^2} {\cal V}_I(\alpha,\ep) \\
&\times&
\sum_{J\neq I}
\bom{T}_I\cdot \bom{T}_J \left(\frac{4\pi\mu^2}{2p_Ip_J}\right)^\ep\;\;,\;\; 
\eeqn
where the indices $I,J$ runs over both final and initial state parton 
and the $p_I\cdot p_J$ dot-products are always positive and
\beqn\nn
{\cal V}_{i}(\alpha,\ep)&=&\bom{T}_i^2\left(\frac1{\ep^2} -\frac{\pi^2}3\right)
+\gamma_i\frac1\ep \\
&+&\gamma_i + K_i(\alpha)+{\cal O}(\ep)\,,
\eeqn
where
\beqn
K_i(\alpha) = K_i - \bom{T}_i^2 \ln^2\alpha 
+ \gamma_i\left(\alpha-1-\ln\alpha\right)
\;\;.
\eeqn
The $K_i$ and $\gamma_i$ constants are defined in the Eq.~(5.90) and
Eq.~(7.28) in Ref.~\cite{Catani:1997vz}.
The $\bom{K}(x,\alpha)$ flavour kernel is given by the followings:
\begin{widetext}
\beqn
\nn
{\bf K}^{a,a'}(x,\alpha) &=& \frac\as{2\pi} \Bigg\{
\overline{K}^{aa'}(x,\alpha) -K^{aa'}_{F.S.}(x)
 -\bom{T}_b\cdot\bom{T}_{a'}\frac1{\bom{T}_{a'}^2} 
\widetilde{K}^{aa'}(x,\alpha)\\
 &&\qquad+\delta^{aa'} \sum_i \bom{T}_i\cdot \bom{T}_a \frac{\gamma_i}{\bom{T}_i^2}
\left[\left(\frac1{1-x}\right)_{1-\alpha}+\alpha \,\delta(1-x)\right]
\Bigg\}\;\;,\qquad\quad
\eeqn
where $K^{aa'}_{F.S.}(x)$ is defined by the factorization scheme. 
In the case of the $\overline{\mathrm{MS}}$ scheme this functions are
identically zero. The flavour functions $\overline{K}^{ab}(x,\alpha)$ are  
\beqn
\nn
\overline{K}^{ab}(x,\alpha) &=& \hat{P}'_{ab}(x)
+P^{ab}_{reg}(x)\ln\frac{\alpha(1-x)}{x} 
+\delta^{ab}\bom{T}_a^2 \delta(1-x) \ln^2\alpha \\\nn
&+&\delta^{ab} \left[\bom{T}_a^2
\left(\frac2{1-x}\ln\frac{1-x}x\right)_+ 
-\delta(1-x)\left(\gamma_a+K_a(\alpha)
  -\frac56 \pi^2\bom{T}_a^2\right)\right]\\
&+&\delta^{ab}\bom{T}_a^2\frac2{1-x}
\left(\ln\frac{\alpha(2-x)}{1+\alpha-x}
  -\ln\frac{2-x}{1-x}\Theta(x < 1-\alpha)\right)\;\;.
\eeqn
The $\widetilde{K}^{ab}(x,\alpha)$ flavour kernels are defined by 
\beqn
\nn
\widetilde{K}^{ab}(x,\alpha) &=& P_{reg}^{ab}(x)\ln\frac{1-x}\alpha
+<\hat{P}_{a,b}(x,\ep=0)>\ln \alpha(x)
-\delta^{ab}\bom{T}_a^2
\left[\left(\frac2{1-x}\ln\frac1{1-x}\right)_{1-\alpha}
  +\frac{\pi^2}3 \delta(1-x)\right]\\
&+&\delta^{ab}\bom{T}_a^2\frac2{1-x}
\left[\ln\frac{1+\alpha-x}{\alpha}
-\ln(2-x)\Theta(x > 1-\alpha)
\right]\;\;,
\eeqn
\end{widetext}
where function $\alpha(x)$ is given by the following
\beqn
\alpha(x) = \min\left\{1,\frac\alpha{1-x}\right\}\;\;.
\eeqn
The definition of the Altarelli-Parisi probabilities
($P^{ab}_{reg}(x)$, $\hat{P}'_{ab}(x)$) and splitting functions 
($<\hat{P}_{a,b}(x,\ep)>$) are given in Appendix C in
Ref.~\cite{Catani:1997vz}. The ``$\beta = 1-\alpha$'' prescription is defined by
\beqn
\int_0^1dx\left(\frac1{1-x}\right)_{\beta} f(x) = 
\int_\beta^1dx\frac{f(x)-f(1)}{1-x}\;\;.
\eeqn

\subsection{Structure of the results}

Once the phase space integrations are carried out, we write the NLO
jet cross section in the following form: \beqn\nn
\label{signjet}
\sigma^{n\rm{jet}}_{AB} &=& \sum_{a,b} \int_0^1 d\eta
\int_0^1d\bar{\eta}
f_{a/A}(\eta, \mu_F^2) f_{b/B}(\bar{\eta}, \mu_F^2) \\
&\times&\hat{\sigma}_{ab}^{n\rm{jet}} \left[p_a, p_b,\as(\mu_R^2),
  \mu_R^2/Q_{HS}^2, \mu_F^2/Q_{HS}^2\right]\;,\; \eeqn where
$f_{i/H}(\eta, \mu_F^2)$ represents the patron distribution function
of the incoming hadron defined at the factorization scale $\mu_F = x_F
Q_{HS}$, $\eta_{a,b}$ is the fraction of the proton momentum carried
by the scattered partons $p_{a,b}$, $Q_{HS}$ is the hard scale that
characterizes the parton scattering which could be $E_T$ of the jet,
jet mass of the event, {\it etc} and $\mu_R = x_R Q_{HS}$ is the
renormalization scale.

In the presented leading- and next-to-leading order results we use the
C/C++ implementation \cite{c-lhpdf} of the LHAPDF library
\cite{Giele:2001mr} with CTEQ6 \cite{Pumplin:2002vw} parton
distribution function and with the corresponding $\as$ expression
which is included in this library.  The CTEQ6 set was fitted using the
two-loop running coupling with $\as(M_{Z^0}) = 0.118$.

To ensure the correctness of the result several check was performed:
(i) the Born level two-, three- and for-jet cross sections were
compared to the prediction of NJETS \cite{Kuijf:1991kn} program and
perfect agreement was found; (ii) singular behaviour of the real and
dipole subtraction terms were checked numerically in randomly chosen
phase space points; (iii) the total NLO corrections was independent of
the $\alpha$ parameters that controls the dipole phase space; (iv) the
program is based on the NLOJET++ program library \cite{c-lhpdf} that
was used other already well tested processes.

In the NLO calculation the difference of the real contributions and
the dipole subtraction terms is still singular but these are
integrable square root singularities. Integrating these singularities 
by simple Monte Carlo integration technique (choosing
random values of the integration variables uniformly) is not efficient
way because the variance of the estimate of the integral is infinite
and we are not able to estimate the statistical error of the integral.
To improve the stability and the convergence of the Monte Carlo
integral the phase space is generated by multi-channel weighted phase
space generator \cite{Kleiss:1994qy}.

\section{\label{sec:result}Results}

In this section we study those jet observables which could be interesting 
in the jet studies at hadron-hadron colliders. We show some examples of 
inclusive three-jet cross sections, energy fraction distributions of the 
jets, event shapes variables and sub-jet rates. To present inclusive jet cross 
section and some related jet observables like energy fraction distribution
we use the Ellis-Soper inclusive $k_\perp$ algorithm \cite{Ellis:1993tq} and the 
midpoint cone algorithm \cite{Blazey:2000qt}. To do sub-jet studies and calculate 
some jet clustering algorithm related event shape variable like flipping 
variable we use the exclusive $k_\perp$ algorithm which was defined by 
Catani {\em et. al.} in Ref.~\cite{Catani:1993hr}. We also present some jet clustering 
independent event shape distribution which are defined in the transverse plane 
(transverse thrust and transverse jet broadening).

\subsection{Three-jet inclusive cross sections}

The most important jet quantities are the inclusive jet cross section
and its transverse distribution. In hadron-hadron collision for jet
studies a variant of the cone algorithms was used. Unfortunately
those algorithms which were used at the first run of the Tevatron are
infrared and/or collinear unsafe or only safe up to a given order in
the perturbative calculation. It is essential that the used jet
algorithm to be infrared safe all order otherwise the perturbative
calculation and the perturbative prediction are unstable. Using an
{\it almost} infrared unsafe jet algorithm we are not able to estimate
the effect of the higher order contributions due to the variation of
the renormalization and factorization scales. The other problem with
this type of the jet algorithms is that they are logarithmically
sensitive to the detector resolution (energy and angular resolution).
That is bad because in the fix order perturbative calculation it is
impossible or hard to consider or simulate precisely these effects 
\cite{Seymour:1998kj}.

In this subsection we use the Ellis-Soper inclusive $k_\perp$
algorithm and the midcone iterative cone algorithm which are infrared
and collinear safe to all orders.  In both cases we consider only those
jets as hard final state jets which have at least $20$GeV transverse
energy and are in the $|\eta| < 2$ pseudo-rapidity interval. The total
jet transverse energy must be larger than $80$GeV.  The
renormalization and factorization scales are characterised by the hard
scale which is the third of the total jet transverse momentum
 \beqn\label{scale-incl}
Q_{HS} = \frac13 \sum_{j=1}^{n_J} p_T^{\mathrm{jet}}(j)\;\;.
\eeqn
All jet contributes to this sum which passed the cuts.

\begin{figure*}[t]
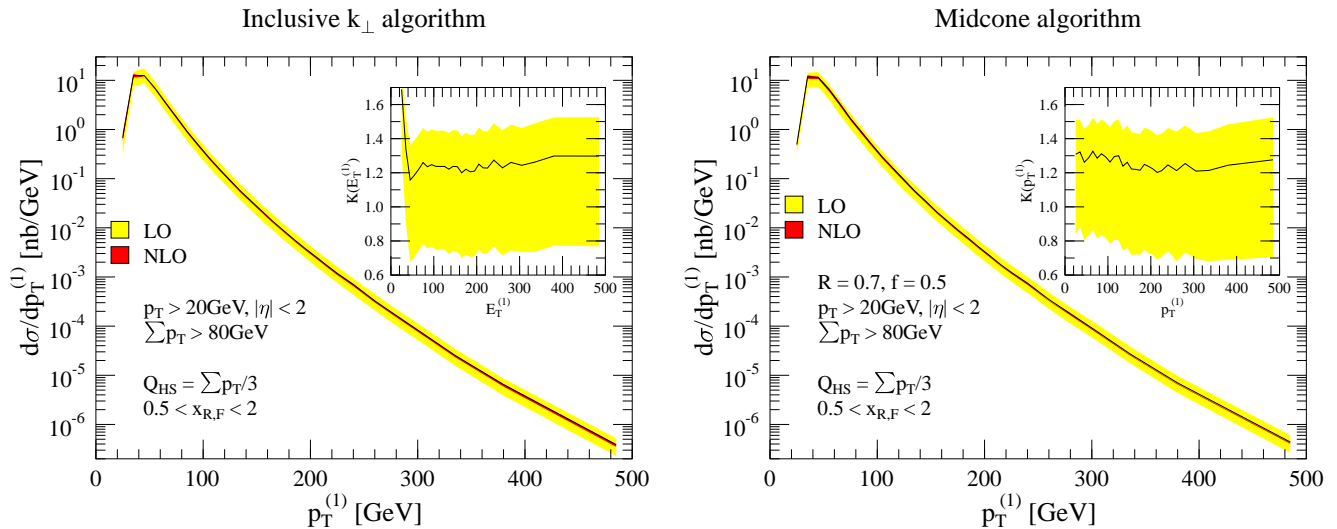
\hfill
  \epsfig{file=fig1.epsi, width = 8.5cm}\hfill \epsfig{file=fig2.epsi,
    width = 8.5cm}\hfill
  \caption{\label{fig:incl} The fix order QCD predictions for the inclusive
    three-jet differential cross sections of the transverse momentum
    of the leading jet obtained using the inclusive $k_\perp$ and
    midcone algorithms.  The bands indicate the theoretical
    uncertainty due to the variation of the scales $x_{R,F}$ between
    $0.5$ and $2$. The grey band is the leading order and the dark
    grey band is the next-to-leading order. The inset figures show the
    $K$ factor and its scale dependence. The solid line represents the
    $x_{R,F} = 1$ scale choice.  }
\end{figure*}
Calculating the transverse energy distribution of the leading jet and
comparing the leading and the next-to-leading order fix order
predictions we can see in the $k_\perp$ case if the scales are set to
$x_{R,F} = 1$ the NLO correction is about $25-30\%$ as shown in
Figure~\ref{fig:incl}. The $K = \sigma^{NLO}/\sigma^{LO}$ factor 
strongly depends on the scale choice but the NLO prediction is
stable. Varying the scales between
$0.5$ and $2$ the scale uncertainty of the NLO prediction is about
$8-10\%$ which is significantly smaller than the leading order
uncertainty which is about $100\%$ or more.

Using the midcone algorithm we got a very similar result. In this
calculation the cone size is $R=0.7$ and the merge/split threshold
parameter is $f=0.5$. In this case the NLO correction is about
$20-30\%$ with the $x_{R,F} = 1$ scales.

In Table~\ref{table:iclusive} the NLO cross sections are tabulated.
Comparing the values we can see that $k_\perp$ algorithm gives higher
cross section in the low $E_T$ region while in the high $E_T$ region
the cone result is higher.  The difference between the two results is
less than $10\%$ almost everywhere.
\begin{table}
  \caption{ \label{table:iclusive} Next-to-leading order results for
    the differential distribution the three jet inclusive cross sections 
    as function of the transverse energy of leading jet. The scales are 
    $x_{R,F} = 1$. The errors are the statistical error of the Monte Carlo 
    integration.
  }
\begin{ruledtabular}
  \begin{tabular}{cll}
    $E_T^{(1)}$  &  \quad$k_\perp$ algorithm & \quad midcone algorithm  \\ 
    GeV     &  \qquad nb/GeV    &  \qquad \quad nb/GeV  \\ 
    \colrule
    $20-30$   &  \res{6.745}{0.240}{-1} &  \res{4.961}{0.387}{-1} \\ 
    $30-40$   &  \res{1.245}{0.009}{1}  &  \res{1.174}{0.019}{1}  \\   
    $40-50$   &  \res{1.252}{0.009}{1}  &  \res{1.148}{0.014}{1}  \\   
    $50-60$   &  \res{6.966}{0.042}{0}  &  \res{6.522}{0.077}{0}  \\   
    $60-70$   &  \res{3.509}{0.020}{0}  &  \res{3.338}{0.034}{0}  \\   
    $70-80$   &  \res{1.799}{0.009}{0}  &  \res{1.662}{0.017}{0}  \\   
    $80-90$   &  \res{9.175}{0.049}{-1} &  \res{9.003}{0.096}{-1} \\   
    $90-100$  &  \res{4.991}{0.026}{-1} &  \res{4.884}{0.054}{-1} \\   
    $100-110$ &  \res{2.755}{0.015}{-1} &  \res{2.713}{0.028}{-1} \\   
    $110-120$ &  \res{1.583}{0.009}{-1} &  \res{1.595}{0.018}{-1} \\   
    $120-130$ &  \res{9.362}{0.057}{-2} &  \res{9.452}{0.011}{-2} \\   
    $130-140$ &  \res{5.554}{0.035}{-2} &  \res{5.607}{0.069}{-2} \\   
    $140-150$ &  \res{3.461}{0.022}{-2} &  \res{3.517}{0.045}{-2} \\   
    $150-160$ &  \res{2.164}{0.015}{-2} &  \res{2.152}{0.030}{-2} \\   
    $160-170$ &  \res{1.350}{0.010}{-2} &  \res{1.398}{0.019}{-2} \\   
    $170-180$ &  \res{8.784}{0.068}{-3} &  \res{9.103}{0.135}{-3} \\   
    $180-190$ &  \res{5.725}{0.049}{-3} &  \res{6.110}{0.099}{-3} \\   
    $190-200$ &  \res{3.806}{0.032}{-3} &  \res{4.014}{0.065}{-3} \\   
    $200-210$ &  \res{2.592}{0.022}{-3} &  \res{2.698}{0.048}{-3} \\   
    $210-220$ &  \res{1.740}{0.016}{-3} &  \res{1.786}{0.034}{-3} \\   
    $220-230$ &  \res{1.185}{0.011}{-3} &  \res{1.243}{0.025}{-3} \\   
    $230-250$ &  \res{7.048}{0.057}{-4} &  \res{7.503}{0.121}{-4} \\   
    $250-270$ &  \res{3.270}{0.036}{-4} &  \res{3.471}{0.080}{-4} \\   
    $270-290$ &  \res{1.646}{0.020}{-4} &  \res{1.797}{0.042}{-4} \\   
    $290-320$ &  \res{7.041}{0.068}{-5} &  \res{7.754}{0.155}{-5} \\   
    $320-350$ &  \res{2.530}{0.031}{-5} &  \res{2.781}{0.057}{-5} \\   
    $350-410$ &  \res{6.471}{0.064}{-6} &  \res{6.993}{0.129}{-6} \\   
    $410-560$ &  \res{3.864}{0.044}{-7} &  \res{4.391}{0.096}{-7} 
  \end{tabular}
  \end{ruledtabular}
\end{table}

\subsection{Topology of the three-jet events}

One can study the structure of the three-jet events. We can define variables
which test certain properties of the jets. These quantities could be any 
azimuthal correlation between the jets or the energy fraction of the three 
leading jets.

The energy fraction of the three leading jets (leading in transverse
energy) was measured by the CDF collaboration
\cite{Abe:1992ui,Brandl:2000ik}. This quantity is defined in centre of
mass frame of the three jets. The jets are ordered and labeled by
energy in the rest frame ($E_1 \, > \, E_2\, >\, E_3$). The energy
fraction variable (usually called Dalitz variables) are given by 
\beqn
X_i = \frac{2E_i}{E_1+E_2+E_3}\;\;, \qquad i = 1,2,3\;\;.
\eeqn 
The variable $X_1$ varies between $2/3$ and $1$, $X_2$ between $1/2$ and
$1$, and $X_3$ between $0$ and $2/3$. There are only two independent
variables since $X_1+X_2+X_3=2$. We use only the first two variables.

In the experimental analysis the iterative cone algorithm was used with an 
additional cut on the jets. It was required that the jets were well separated.
This additional cut ensures that the cross section is calculable order by order
in perturbation theory but the jet algorithm is still not fully infrared 
safe, it is almost infrared unsafe \cite{Seymour:1998kj}.

To get stable theoretical predictions in our calculations we use the 
inclusive $k_\perp$ and midcone jet algorithms to resolve jets with the same 
cuts what was used in Ref.~\cite{Brandl:2000ik}. All jets are required to have at
least $20$GeV transverse energy and they must lay in the $|\eta| < 2$ 
pseudo-rapidity window. We calculate the normalised differential distributions 
of the $X_1$ and $X_2$ variables and the double differential distribution of 
these variables. In these calculations the renormalization and factorization 
scales are set to the third of the total jet transverse energy 
Eq.~(\ref{scale-incl}).

\begin{figure}
  \epsfig{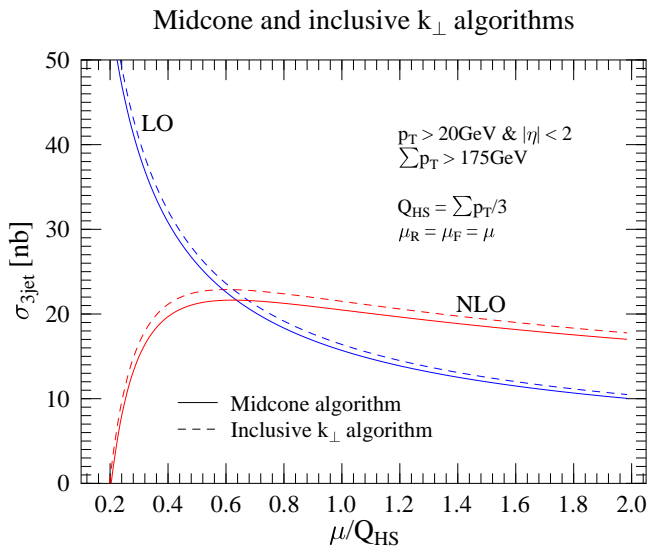}
  \caption{\label{fig:scale} The scale dependence of the total 3-jet cross 
    sections. 
  }
\end{figure}
The distributions are normalised by the total 3-jet cross section plotted in 
Figure~\ref{fig:scale}. We can see the NLO correction reduced the scale 
dependence of perturbative prediction. The scale uncertainty of the NLO result 
is about $10-12\%$ varying the scales on the $0.6 < x_{R,F} < 2$ range while 
the LO uncertainty is much higher $38-40\%$. The corrections with the 
$x_{R,F} = 1$ scale choice are about $30\%$.  

\begin{figure*}
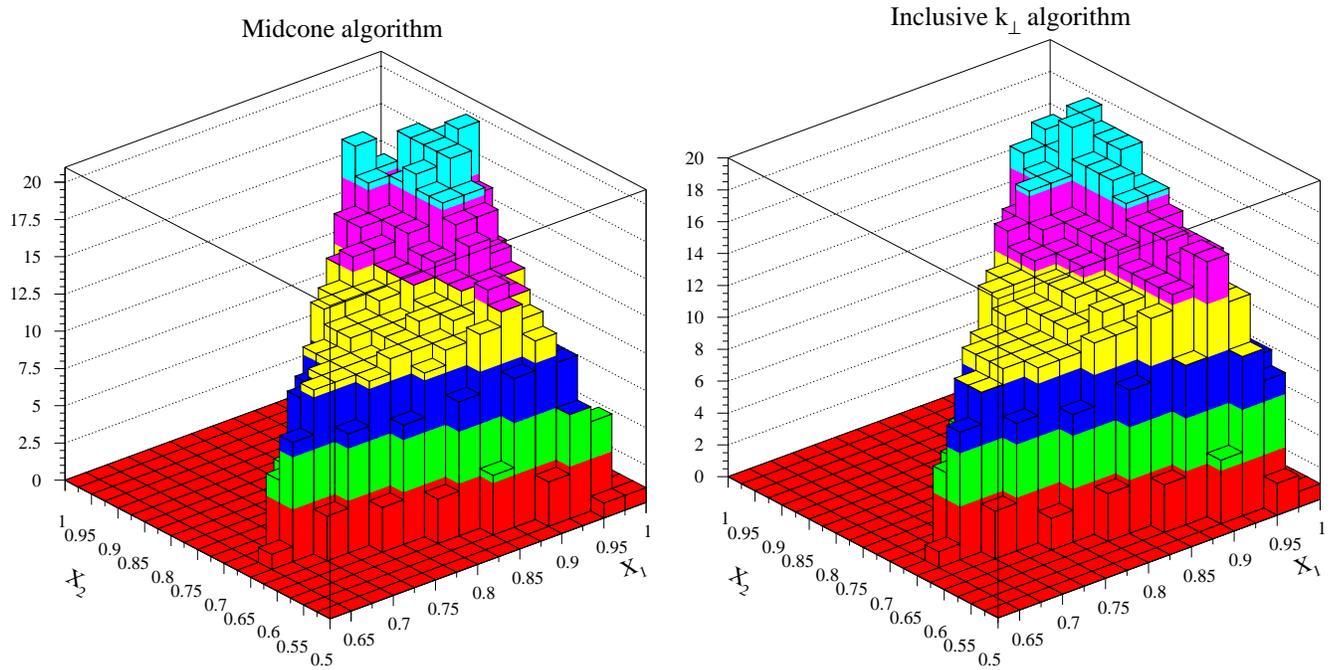
\hfill
  \epsfig{file=fig8.epsi, width = 8.5cm}\hfill
  \epsfig{file=fig9.epsi, width = 8.5cm}\hfill
  \caption{\label{fig:X1X2} Next-to-leading order perturbative prediction 
    for normalised double differential distribution ($1/\sigma d\sigma/dX_1dX_2$)
    of the energy fraction variables $X_1$ and $X_2$ using the midcone and 
    inclusive $k_{\perp}$ algorithm. 
  }
\end{figure*}
The double differential distributions of the variables $X_1$ and $X_2$ are 
plotted in Figures~\ref{fig:X1X2}. The phase space would populate the 
available $X_1$, $X_2$ region uniformly. Deviation from the
uniform distribution shows the effect of the QCD dynamics. 

Taking the double differential distribution and projecting on either axis the 
distributions of variables $X_1$ and $X_2$ can be obtained.
\begin{figure}
  \epsfig{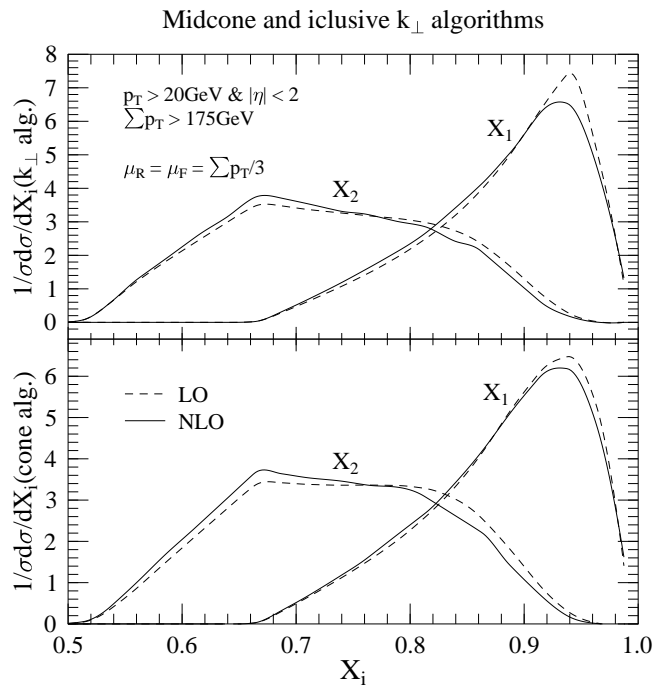}
  \caption{\label{fig:Xi} The energy fraction distribution of the leading 
    ($X_1$) and second leading ($X_2$) jets. The upper figure is result with
    the inclusive $k_\perp$ algorithm and the lower figures shows the midcone 
    result.
  }
\end{figure}
In Figure~\ref{fig:Xi} the differential distributions of the energy fractions 
are plotted in the inclusive $k_\perp$ and midcone algorithm cases.  
 
Comparing the leading order result to the next-to-leading order result we can 
see that the NLO distribution of the leading jet is less peaked than the LO 
but in the case of the $X_2$ distribution the NLO result is more peaked
around $0.65$. The NLO correction of the distributions is small.

Comparing the two algorithms to each other we can say that the distributions 
are very similar and there is not too much difference. The $X_1$ distribution 
of the inclusive $k_\perp$ algorithm is a little bit more peaked and the $X_2$ 
distribution is a little bit flatter than in the midcone case. The 
corresponding peaks are in the same positions.

\subsection{Three-jet event shapes}
 
We can distinguish two type of the event shapes. One can calculate and
measure the event shapes which are associated to a jet algorithm.
Generally we can say that this type of variables gives information
about any geometrical property of the jets in a $n$-jet event.  Or we
can define event shape variables which measure any geometrical
property of the event {\it e.g} thrust or jet bordering.

Using the exclusive $k_\perp$ algorithm there is a natural way to
introduce event shape variables. The prescription of this algorithm
introduces a stopping parameter $d_{cut}$ (resolution variable), that
defines the hard scale of the process and separates the event into a
hard scattering part and a low-$p_T$ part ({\it beam-jet}).

Defining the $d_{cut}$ resolution variable event-by-event we can
calculate the differential distribution of the $E_{tn}$ event shape
variable. Denoting by $d_{m}$ the smallest resolution variable when
the event has $m$ hard final state jets the $E_{tn}$ event shape
variable is defined by
\beqn 
E_{tn}^2 = \max_{m \ge n} \{d_m\}\;\;.
\eeqn

\begin{figure}
  \epsfig{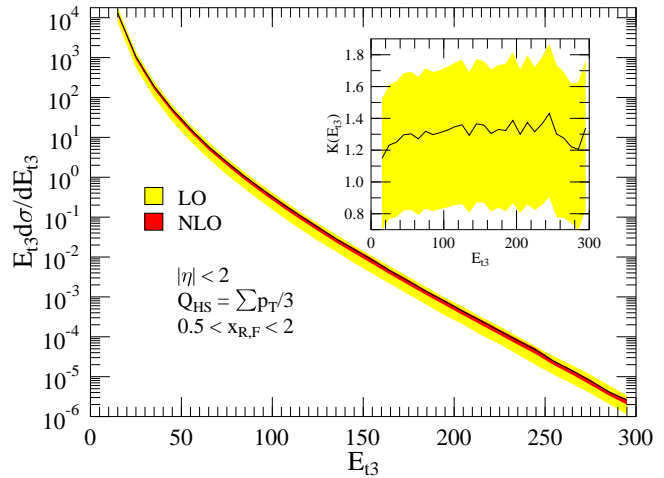}
  \caption{\label{fig:et3} Distribution of the $E_{t3}$ event shape 
    variable at LO an NLO level. The bands indicate the theoretical 
    uncertainty due to the variation of the scales 
    $x_{R,F}$ between $0.5$ and $2$. The grey band is the leading order and 
    the black band is the next-to-leading order. The inset figure shows the 
    $K$ factor and its scale dependence. The solid line represents the 
    $x_{R,F} = 1$ scale choice.
  }
\end{figure}
We plotted the differential distribution of the $E_{t3}$ event shape variable
in Figure~\ref{fig:et3}. In this calculation the $|\eta| < 2$ pseudo-rapidity 
cut was applied for every hard final state jet. The renormalization and 
factorization scales were given by $\mu_{R,F} = x_{R,F} Q_{HS}$ where the hard 
scale is the average transverse momentum of the three jets
\beqn
Q_{HS} = \frac13 \sum_{j=1}^3 p_T^{\mathrm{jet}}(j) \;\;.
\eeqn
Comparing the next-to-leading order result to the leading order result we see 
that the LO prediction strongly depends on the scales. Varying the scale 
parameters $x_{R,F}$ between $0.5$ and $2$ the scale uncertainty of the NLO 
result is about $15\%$ in contrast to the LO scale uncertainty which is about 
$80\%$. Setting the scales to $x_{R,F} = 1$ we see that the NLO correction is 
about $20-35\%$.

One can define event shapes on the transverse plane. Important example is the 
transverse thrust which is defined by
\beqn
T_\perp = \max_{\vec n} 
\frac{\sum_{i \in C_N} |\vec p_{\perp,i}\cdot \vec n|}
{\sum_{i \in C_N} |\vec p_{\perp,i}|}\;\;,
\eeqn
where $\vec p_{\perp,i}$ is the transverse component of the parton
(hadron) momentum. The unit vector that $\vec n$ maximises the ratios
of the sums is usually called thrust axis. In these sum only those
particles are counted which are fulfil all selection criteria $C_N$.
This selection must be infrared safe. In this calculation we required
that the pseudo-rapidity of those particles which contribute to
$T_\perp$ must be in the $[-1.1,1.1]$ pseudo-rapidity window thus we have
\beqn
C_N = \{i\;\;:\quad |\eta_i| < 1.1\;\;, \quad i = 1,\dots,N\}\;\;,
\eeqn
where $N$ is the number of particles in the event. This definition of
the thrust for hadron colliders gives an infrared safe longitudinally
boost invariant quantity.

The jet broadening $B_\perp$ is an associated quantity to the
thrust. Using the thrust axis we can define $B_\perp$ on the transverse plane
by the following formulae
\beqn
B_\perp = \frac{\sum_{i \in C_N} |\vec p_{\perp,i}\times \vec n|}
{2\sum_{i \in C_N} |\vec p_{\perp,i}|}\;\;.
\eeqn

\begin{figure}
  \epsfig{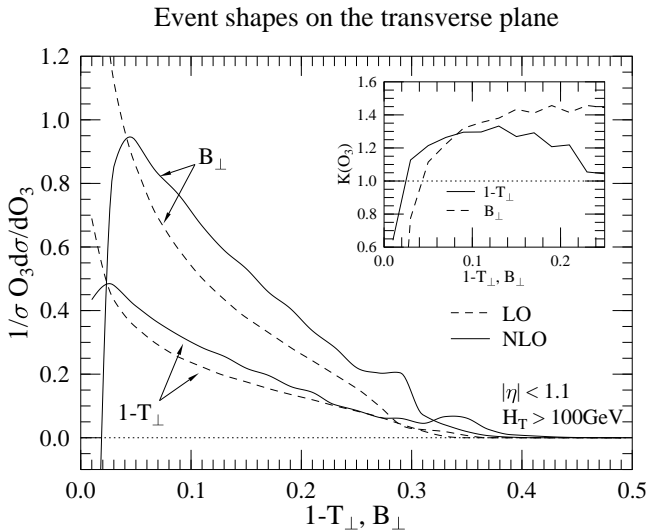}
  \caption{\label{fig:thrust-dist} Distribution of $1-T_\perp$ and $B_\perp$ 
    event shape variable at LO ({\it dashed line}) and NLO ({\it solid line}) level. 
    The scales were set to $\mu_{R,F} = H_T$. The inset figure shows the 
    $K$-factors.
  }
\end{figure}

In Figure~\ref{fig:thrust-dist} the differential distributions of the 
transverse thrust and the transverse jet broadening have been plotted 
at LO and NLO level. In this calculation the total transverse energy 
is larger than $100$GeV
\beqn
H_T \equiv \sum_{i\in C_N} E_{T,i} > 100\mathrm{GeV}\;\;.
\eeqn
The perturbative results logarithmically divergent at the edge of the phase 
space. This divergence occurs at $T_\perp=1$ and $T=2/3$ in the thrust 
distribution and at $B_\perp=0$ and $B=1/3$ in the jet broadening distribution.
Setting the scale to the total transverse energy the NLO order correction 
in the middle of the distribution where the effect of the logarithm is small
is $30-35\%$ for the thrust and  $20-45\%$ for jet broadening as it is shown 
in the inset figure. It is necessary to do all order resummation of the 
leading and next-to-leading logarithms to be able to do quantitative 
comparison of the theory to the data.

One can define the average value of the event shape variables as
function of the total transverse energy. This quantity is given bay
the following ratio: 
\beqn
\langle O_3 \rangle_{\delta} = \left(\frac{d\sigma}{dH_T}\right)^{-1}
\int_{\delta}^1 dO_3 O_3\frac{d\sigma}{dO_3dH_T}\;\;,
\eeqn
where $O_3$ could be either $1-T_\perp$ or $B_\perp$.  

\begin{figure}
  \epsfig{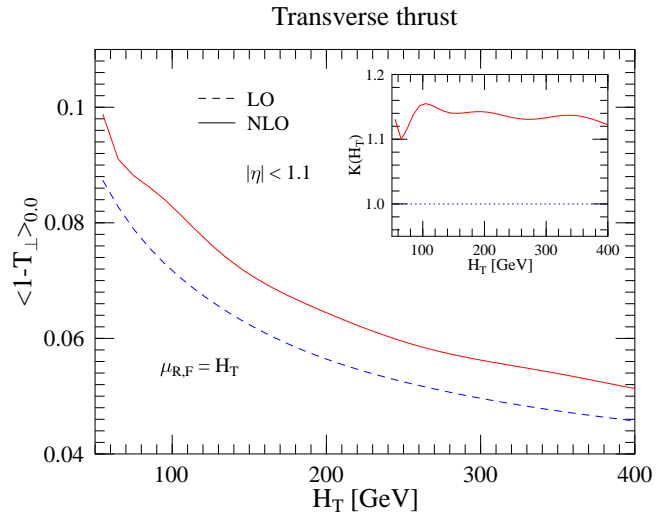}
  \caption{\label{fig:avthrust} Average value of the $1-T_\perp$
    event shape variable at $\delta=0$ at LO ({\it dashed line}) and
    NLO ({\it solid line}) level. The scales were set to $\mu_{R,F} =
    H_T$. The inset figure shows the $K$-factor.
  }
\end{figure}

Figure~\ref{fig:avthrust} shows the average value of the transverse
thrust. With the  $\mu_{R,F} = H_T$ scale choice it was found that the
NLO correction is small about $15\%$. The NLO correction doesn't
change the shape of the distribution as it is shown by the K-factor
plot. This result indicate that the NLO fix order prediction might
describe the data itself with good accuracy.    

\begin{figure}
  \epsfig{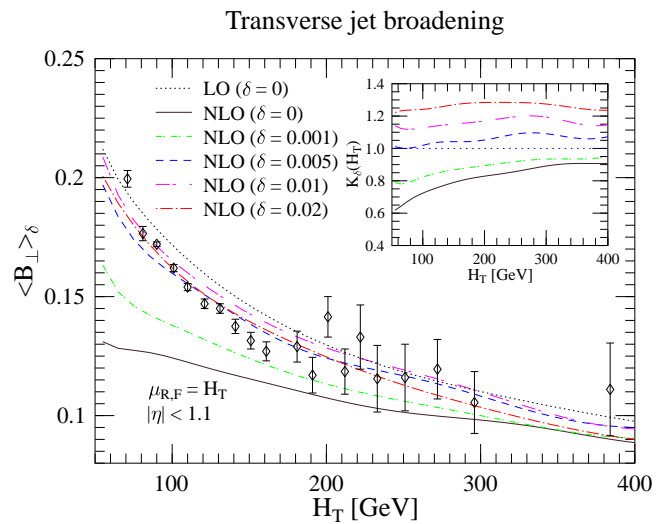}
  \caption{\label{fig:avjetbr} Average value of the $B_\perp$
    event shape variable at $\delta=0$ at LO and NLO level for $\delta
    = 0, 0.001,0.005,0.01,0.02$. The scales were set to $\mu_{R,F} =
    H_T$. The markers ({\it diamonds}) and the error bars represent
    the CDF data \cite{Abe:1991dm} and its statistical error. The
    inset figure shows the $K$-factors.}
\end{figure}

Figure~\ref{fig:avjetbr} shows the average value of the transverse
jet broadening. The CDF collaboration \cite{Abe:1991dm} measured this
distribution with $\delta=0$ choice and the data was compared to the
leading order QCD prediction. Comparing the data to the NLO prediction
the difference between the data and the theory is huge. This can
happen because the contribution of the large logarithms from the small
$B_\perp$ region is very large and in the NLO case it causes a huge
negative effect. It is clear that the fix order calculation is unable
to describe the data itself. The resummation of the large logarithms is
important \cite{Banfi:2003je}.

As shown in Fig.~\ref{fig:thrust-dist} the large negative
contributions come form the small $B_\perp$ region ($B_\perp < 0.02$).
Resuming the leading and the next-to-leading logarithms the $B_\perp
d\sigma/dB_\perp$ distribution is positive definite and we can expect
that it is reasonable small in the small $B_\perp$ region. On the other
hand Fig.~\ref{fig:thrust-dist} indicates that the NLO fix order
calculation is stable prediction in the large ($B_\perp > 0.2$) region.
Assuming that the all order contribution of the $B_\perp < 0.02$
region to the average value is small and changing the $\delta$
parameter we can make a better comparison of the data and the NLO
prediction. In Fig.~\ref{fig:avjetbr} we can see that the $\langle
B_\perp\rangle_{0.0}$ and the $\langle B_\perp\rangle_{0.001}$ NLO
results suffer on the effect of the large logarithm but changing the
$\delta$ parameter between $0.005$ and $0.02$ the NLO prediction is
stable and  hardly depends on the $\delta$ parameter and the agreement
with the data is much better. The inset figure shows the $K_\delta =
\langle B_\perp\rangle_\delta^{NLO}/\langle
B_\perp\rangle_\delta^{LO}$ factor. For $0.005 < \delta < 0.02$ the
NLO correction is positive, nearly constant and it is maximum
$30\%$. 

Of course this is not a precise analysis and without the resummation
of the leading and next-to-leading logarithms the quantitative
comparison of the current data and the theory is impossible but it can
help to understand the huge discrepancy between the data and the NLO
prediction. It also indicate that the NLO prediction can
describe the data itself very well if $\delta > 0.005$.

\subsection{Sub-jet fractions}

In this subsection we study the sub-jet multiplicities of the events.
We use the exclusive $k_\perp$ algorithm \cite{Catani:1993hr}. The
jets are defined in a two-step clustering procedure. The first step of
the algorithm identifies the low-$p_T$ scattering fragments and includes
them in the beam jets, thus factorizing the hard scattering
subprocess. The scale of the hard final state jets is $E_{cut}$ the
stopping parameter of this clustering step. Defining the sub-jet
resolution variable $y_{cut} = Q_0/E_{cut}$ the second step of the
algorithm resolves the sub-jet structure of the hard final state jets.

The $n$ sub-jet rate is defined by the ratio of the $n$ sub-jet cross section 
and the total cross section as 
\beqn
R_n(E_{cut}, y_{cut}) = \frac{\sigma_n(E_{cut}, y_{cut})}{\sigma_{tot}(E_{cut})} \;\;,
\eeqn
where the $\sigma$ is the total cross section defined by the sum of the 
$n$-jet cross section at the $E_{cut}$ scale
\beqn\label{tot-xsec}
\sigma_{tot}(E_{cut}) = \sum_{n=2}^\infty \sigma_n(E_{cut})
= \int_{E_{cut}}^\infty dE_{t2} \frac{d\sigma}{dE_{t2}}\;\;.
\eeqn
The $E_{t2}$ variable is the smallest value of the $E_{cut}$ resolution 
variable where the event has two jets. Using the similar event shape 
variables $y_n$ and $y_{n+1}$ of the second step we can define the exclusive 
sub-jet cross section at a given $E_{cut}$ scale by
\beqn\nn
\sigma_n(E_{cut},y_{cut}) &=& \int_{E_{cut}}^\infty dE_{t2}
\left[
 \int_{y_{cut}}^\infty dy_n \frac{d\sigma}{dE_{t2} dy_n} \right.
\\
&&\quad - 
\left.
\int_{y_{cut}}^\infty dy_{n+1}
 \frac{d\sigma}{dE_{t2} dy_{n+1}} 
\right]\;\;.\;\;
\eeqn

In the fix order perturbation calculation we can calculate the $3$-sub-jet 
ratio at NLO level ($R_3^{NLO}(E_{cut}, y_{cut})$) and the $4$-sub-jet ratio 
at LO level ($R_4^{LO}(E_{cut}, y_{cut})$). From the definition 
of the $n$-sub-jet ratios we have the normalisation condition
\beqn
1 = \sum_{i=2}^\infty R_n(E_{cut}, y_{cut})\;\;.
\eeqn
Using this relation the 2-sub-jet ration can be obtained at 
next-to-next-to-leading order level (NNLO)
\beqn\nn
&&R_2^{NNLO}(E_{cut}, y_{cut}) =  1-R_3^{NLO}(E_{cut}, y_{cut})\\
&&\qquad\qquad -R_4^{LO}(E_{cut}, y_{cut}) + {\cal O}(\as^3)\;\;.\;\;
\eeqn 

Analogously to the $e^+e^-$-annihilation we can define the sub-jet multiplicity
\beqn
N(E_{cut}, y_{cut}) = \sum_{n=2}^\infty n R_n(E_{cut}, y_{cut})\;\;,
\eeqn
and at up to second order
\beqn\nn
&&N(E_{cut}, y_{cut}) = 2 + R_3^{NLO}(E_{cut}, y_{cut}) \\
&&\qquad\qquad + 2R_4^{LO}(E_{cut}, y_{cut}) +{\cal O}(\as^3)\;\;.
\eeqn

The fix order perturbative predictions depend on the unphysical 
renormalization and factorization scales. In these calculations the scales 
are given by event-by-event basis. We use the invariant mass of the two hard 
final state jets found at $E_{t2}$ scale as renormalization and factorization
scale
\beqn
\mu^2_{R,F} = (p_1 + p_2)^2\;\;.
\eeqn

\begin{figure*}[t]
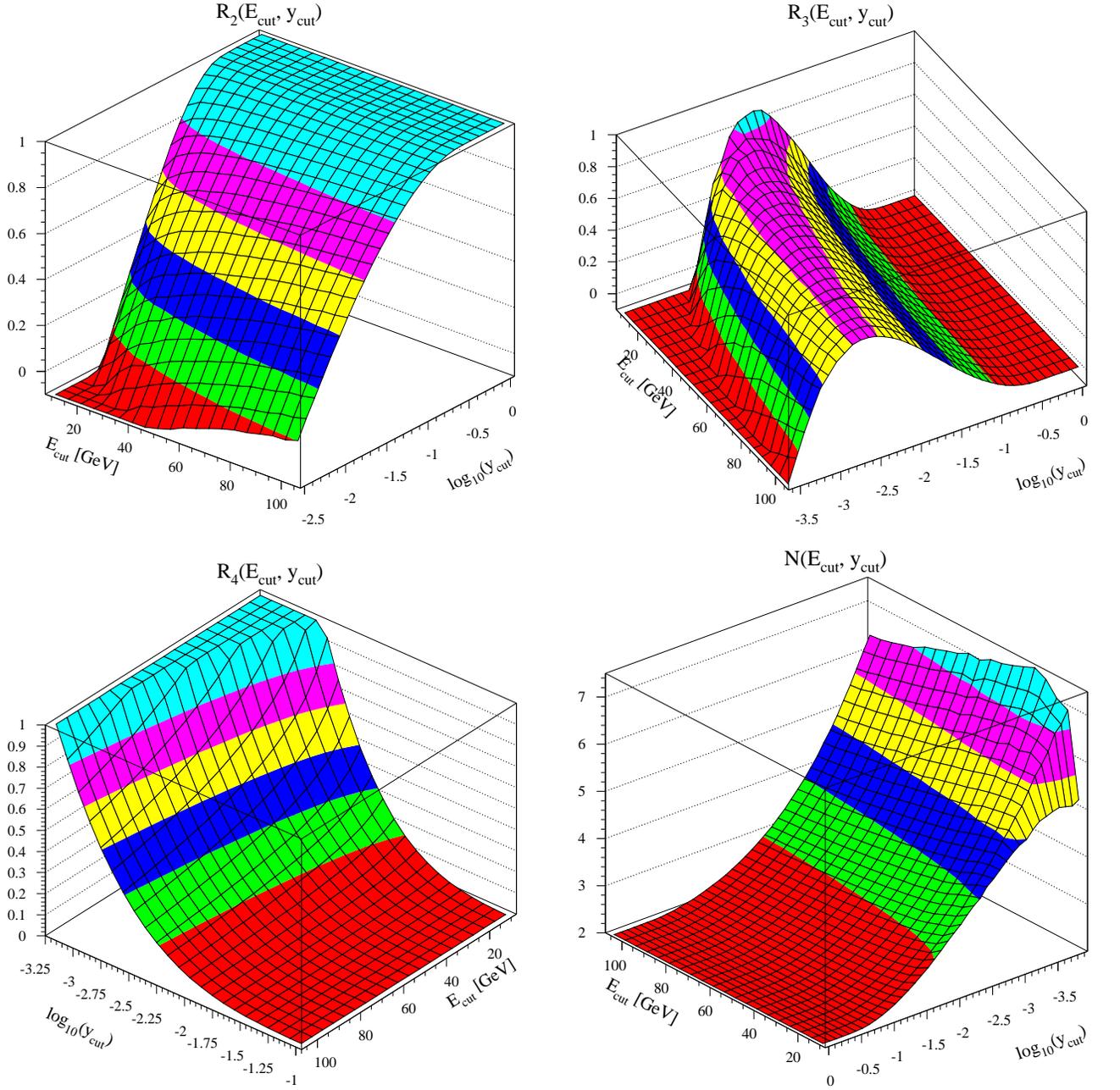
\hfill
  \epsfig{file=R2.epsi, width = 8.cm}\hfill
  \epsfig{file=R3.epsi, width = 8.cm}\\\hfill
  ~\\\hfill 
  \epsfig{file=R4.epsi, width = 8.cm}\hfill
  \epsfig{file=N.epsi, width = 8.cm}\hfill
\caption{\label{fig:Rn} The fix order QCD prediction for the two-, three- 
  and four-jet sub-jet rates and for the sub-jet multiplicity. The two-jet 
  rate is calculated at NNLO, the three-jet rate is calculated at NLO and 
  the four-jet at LO level.
}
\end{figure*}
Fig.\ref{fig:Rn} shows the sub-jet ratios and the sub-jet
multiplicity. The result is given as function of $E_{cut}$ parameter
and the logarithm based ten of the sub-jet resolution variable. The
distributions are plotted on the $10 < E_{cut} < 110\ \mathrm{GeV}$
and $-3.9 < \log_{10}(y_{cut}) < 0$ region.  We can see the sub-jet
ratios are strongly dominated by the large logarithms in the low
$E_{cut}$ and $y_{cut}$ regions. To be able to do any quantitative
comparison to the experimental data in this region it is necessary to sum
at least the leading- and next-to-leading logarithms at all order and
match it to the fix order calculation.

\section{\label{sec:conclusion}Conclusion}

This paper dealt with the next-to-leading order calculation of
three-jet observables in hadron-hadron collision. We gave a
modification of the original dipole method that made possible to
construct a Monte Carlo program for this process. We find this
modified dipole subtraction method was very useful from the point of the
numerical calculation. The introduced phase space cut in the dipole
term could reduce the evaluation time of subtraction terms and
increase the numerical stability of the Monte Carlo integral by
reducing the probability of the missed binning.

We calculated the transverse momentum distribution of the leading jet
of the three-jet inclusive cross section using the inclusive $k_\perp$
an midcone algorithms. We found that the NLO correction can stabilize
the theoretical prediction by reducing the renormalization and
factorization scale dependences significantly. We also calculated the
energy fraction distribution of the three-jet inclusive cross section.
Comparing the leading order result to the next-to-leading order result
we found that the NLO corrections are negligible in these analysises.

We defined event shape variables in the transverse plane. We
calculated the differential distribution of the transverse thrust and
transverse jet broadening and the transverse energy dependence of the
average values. We found that the size of the NLO correction is not too
large and acceptable but at the edge of the available phase space the
result suffers on the contribution of large logarithm.  In this region
a next-to-leading logarithmic approximation (NLL) matched with the fix
order prediction can give reasonable result. The average value of the
transverse jet broadening was compared to the CDF data and we found
that the comparison of the current data and the theory without NLL
resummation is impossible. Using a rough approximation in the small
$B_\perp$ region we were able to do some qualitative comparison
between data and the NLO QCD prediction and we found reasonable good
agreement. 

We also calculated the subjet rates and the subjet multiplicity using
the exclusive $k_\perp$ algorithm. These quantities were defined in
that way to mimic $e^+e^-$ annihilation like situation at hadron
colliders. The result is very similar to what we have in $e^+e^-$
annihilation. The subjet ratios are strongly dominated by the
final-state large logarithm. Upgrading the fix order calculation with
next-to-leading logarithmic approximation this quantity may provide a
sensitive measurement of the strong coupling at hadron colliders.

Recently two general methods have been developed by Banfi, Salam and
Zanderighi \cite{Banfi:2003je} and by Bonciani, Catani, Mangano and
Nason \cite{Bonciani:2003nt} to carry out the next-to-leading
logarithmic resummation for any observable. These methods can help us to
improve our fix order calculation and to be able to make more precise
comparison between the data and the theory.

\section{Acknowledgement}

I wish to thank Zolt\'an Tr\'ocs\'anyi for useful discussions and
suggestions, as well as for collaboration on various elements of this
project.  This work was supported in part by US Department of Energy,
contract DE-FG0396ER40969, by Research Training Network `Particle
Physics Phenomenology at High Energy Colliders', contract
HPRN-CT-2000-00149 as well as by the Hungarian Scientific Research
Fund grant OTKA T-038240. I also thank for Institute for Particle
Physics Phenomenology (IPPP) at University of Durham and for Fermi
National Laboratory to make me possible to use their computer
facilities.

\bibliography{prd}
\end{document}